# Unpacking Graduate Students' Learning Experience with Generative AI Teaching Assistant in A Quantitative Methodology Course

Zhanxin Hao[1], Haifeng Luo[2], Yongyi Chen[1], Yu Zhang*[1]

[1] Institute of Education, Tsinghua University

[2] Beijing Academy of Educational Sciences

*Abstract*

**Background:** This study was motivated by the increasing integration of generative AI (GenAI) tools in educational settings and the need to understand students' interaction patterns and learning experiences with these technologies. Existing knowledge on how students engage with AI, particularly across different academic backgrounds, remains limited.

**Objectives:** The research aimed to explore students' learning experiences by analysing their learning behaviours, as well as their attitudes and perceptions, when interacting with an AI teaching assistant.

**Methods:** The study was conducted in an Advanced Quantitative Research Methods course involving 20 graduate students. During the 10-week course, student inquiries made to the AI were recorded and coded using Bloom's taxonomy and the CLEAR framework. A series of independent sample t-tests and poisson regression analyses were employed to analyse the characteristics of different questions asked by students with different backgrounds. Post-course interviews were conducted with 10 students to gain deeper insights into their perceptions.

---

[1] Correspondence concerning this article should be addressed to Professor Yu Zhang, 4th Floor, Wennan Building, Tsinghua University, Haidian District, Beijing, 100084. Email: zhangyu2011@tsinghua.edu.cn

**Results and Conclusions:** The findings revealed a U-shaped pattern in students' use of the AI assistant, with higher usage at the beginning and towards the end of the course, and a decrease in usage during the middle weeks. Most questions posed to the AI focused on knowledge and comprehension levels, with fewer questions involving deeper cognitive thinking. Students with a weaker mathematical foundation used the AI assistant more frequently, though their inquiries tended to lack explicit and logical structure compared to those with a strong mathematical foundation, who engaged less with the tool. These patterns suggest the need for targeted guidance to optimise the effectiveness of AI tools for students with varying levels of academic proficiency.

# 1. Introduction

The integration of Generative Artificial Intelligence (GenAI) in education has opened new avenues for personalised and adaptive learning. Increasingly, students are relying more and more on GenAI for information searching and assisting self-learning (Yusuf et al., 2024; Pesovski et al., 2024; Chan & Hu, 2023). GenAI tools have been rapidly incorporated into educational systems to help students in various fields (Baidoo-Anu & Ansah, 2023). For example, in medical education, AI tools like ChatGPT simulate real-life cases and offer immediate feedback on diagnostic and treatment decisions, enriching the learning experience (Sallam et al., 2023; Šedlbauer et al., 2024). Similarly, in programming education, AI assistants such as GitHub Copilot support students by offering code suggestions, error detection, and automatic code generation, helping them write more efficient code with less effort (Denny et al., 2022b; Yilmaz & Yilmaz, 2023). Additionally, tools like KNUST-bot, integrated into a multimedia programming course at Kwame Nkrumah University of Science and Technology, provide further support (Essel et al., 2022). Language learning also widely benefits from AI applications. A wide range of language-learning tools has emerged, such as Duolingo, AI-KAKU, designed to assist learners in improving their speaking, writing, reading, and other language skills (Gayed et al., 2022).

    GenAI tools hold unique advantages in supporting student learning. First, GenAI systems possess general intelligence, allowing them to efficiently search for and retrieve relevant knowledge and information. This provides students with rapid access to relevant content, significantly streamlining the search process and aiding their study efforts (Hmoud et al., 2024; Bilquise et al., 2023). Additionally, GenAI's capacity for generalisation and deep reasoning further supports personalised learning by adapting to individual student needs and promoting deeper cognitive engagement. These capabilities make GenAI tools powerful for fostering both broad and targeted educational outcomes (Wood & Moss, 2024; Karpouzis et al., 2024). For instance, Khan Academy has integrated GPT-4 into its platform through Khanmigo, a virtual tutor that provides personalised learning paths, step-by-step explanations, and customised practice exercises (Chen et al., 2023). Moreover, GenAI's advanced language understanding, and generation capabilities allow for natural, fluent communication between students and the AI, creating more personalised and interactive learning experiences. Real-time dialogue fosters deeper conversations,

encouraging students to explore topics in greater depth and promoting a more reflective learning process (Strzelecki, 2023; Guo et al., 2023; Liang et al., 2023).

The interactive dialogue between GenAI tools and students aligns closely with the Socratic method, which emphasises learning through questioning and reflection. Socrates' dialogical method embodies the educational principle of fostering knowledge through interaction, aiming to cultivate individuals who can engage in social life thoughtfully and rationally. Neo-Kantian philosopher Leonard Nelson (1949) argues that the uniqueness of Socratic dialogue lies in its ability to guide learners to think and live like philosophers, promoting an understanding of the essence of things through active exploration, questioning, evidence gathering, and reasoning. Warriner and Anderson (2016) emphasised that discourse analysis, particularly of student speech, is essential for understanding how students learn. In the era of GenAI-assisted learning, the dialogue between students and AI has become a central element of the learning process. Analysing students' verbal exchanges can reveal their cognitive strategies, helping educators better understand how students think, learn, and engage with academic content (Gee, 2014).

The current study explores students' interactions with a GenAI teaching assistant in an authentic higher education context using a mixed research approach. The study focuses on the analysis of students' questions to GenAI, including the timing, types, and nature of their inquiries, as well as differences in questioning behaviours among different students. Questionnaires and interviews were also used to investigate students' learning experiences of using GenAI teaching assistants and their understanding of the effectiveness of integrating GenAI teaching assistants into higher education.

**2. Literature review**

**2.1 GenAI in education**

A substantial body of research exists in fields such as computer science and educational technology, focusing on the design of technology applications in education as well as the evaluation of the effectiveness of these applications. This extensive body of work reflects the longstanding and broad research interest in this topic among researchers. The earliest use of AI in education dates back to the development of PLATO in the 1960s at the University of Illinois (Bitzer et al., 1961; Woolley, 1994) and the SCHOLAR platform in the 1970s (Carbonell, 1970), which marked the rise of Intelligent Tutoring Systems. In recent years, ITS has advanced significantly due to the integration of deep learning techniques. Deep neural networks, particularly in knowledge tracing—a critical component of ITS—have enhanced the ability to model student learning and predict future performance based on prior interactions (LeCun et al., 2015; Piech et al., 2015). Innovations such as Dynamic Key-Value Memory Networks, which incorporate side information like question relationships, have further improved ITS predictive capabilities (Zhang et al., 2017). Research has demonstrated the effectiveness of ITS in educational settings. ITS consistently outperformed traditional classroom teaching and non-ITS computer-based learning methods. Nesbit et al. (2014) reported significant advantages of ITS over teacher-led and non-ITS digital instruction. Similarly, Ma et al. (2014) and Pane et al. (2014) claimed that ITS are effective tools for enhancing learning outcomes. However, Steenbergen-Hu and Cooper (2014) found that while ITS had a moderate positive impact on college students' academic performance, it was less effective than human tutoring.

While these advancements, particularly through deep learning techniques, have greatly enhanced the capabilities of ITS, traditional ITS models still face significant limitations. Specifically, early traditional Intelligent Tutoring Systems (ITSs) typically rely on predefined rules and limited databases to provide personalised learning support (Mousavinasab et al., 2018; VanLehn, 2011). These systems are based on fixed instructional models and use pre-programmed interaction pathways to assess students' understanding and deliver corresponding feedback. Traditional ITSs can predict students' understanding by analysing their errors and response times, providing corresponding hints or additional exercises through predetermined feedback pathways (Corbett et al., 1997). However, the feedback and guidance offered by these systems are based on a limited set of predefined options, which lack flexibility and adaptability (Conati, 2009). As a result, these systems are often less effective in meeting the complex and dynamic needs of students in more varied learning scenarios (Rosenberg, 1987). Compared to ITSs, GenAI tools bring significant innovations. First, they possess vast databases, enabling them to handle a wide range of knowledge domains and diverse student queries. In addition, GenAI can generate and adapt feedback in real-time, offering more targeted and personalised responses compared to traditional ITSs. This context-aware and dynamic interaction allows students to pose more complex and open-ended questions, thereby enhancing engagement and interactivity in the learning process.

**2.2 Existing research on GenAI integration in teaching and learning**

With the release and development of ChatGPT, many researchers have begun exploring the integration of GenAI in education. One of the more widely explored areas is its application in the language learning domain (e.g., Song & Song, 2023; Tai & Chen, 2024; Du & Daniel, 2024; Meyer et al., 2024). Some researchers are investigating the use of GenAI in more general settings. For example, Essel et al. (2024) examined the effects of using ChatGPT by undergraduates to complete classroom tasks, examining its impact on students' cognitive skills. Current research suggests that the potential of GenAI is not limited to improving specific knowledge acquisition or skill development, such as correcting English pronunciation or enhancing fluency; it also has positive effects on students' non-cognitive dimensions.

The use of AI teaching assistants has been shown to improve academic performance across various educational levels when compared to traditional instructor-only support. For example, Tai & Chen (2024) explored the effects of a GenAI chatbot called CoolEBot on assisting elementary school English learners with their speaking abilities. Pre- and post-test comparisons revealed that the group using the chatbot showed significantly higher improvements in speaking ability compared to the non-using group. In another study focusing on writing, Chen and Pan (2022) demonstrated that GenAI tools can enhance students' writing proficiency, particularly in grammar and vocabulary, with significant gains observed among intermediate and introductory English learners. The interactive nature of GenAI creates a responsive and engaging learning experience, improving both foundational understanding and independent exploration of additional resources. Essel et al. (2022) highlighted the challenge of providing timely feedback in traditional settings, noting that AI-powered teaching assistants effectively address this issue, leading to higher student satisfaction with chatbot use.

From a non-cognitive perspective, existing literature reveals that GenAI tools hold significant potential to enhance students' learning motivation, psychological comfort, and self-regulated learning abilities (Du & Daniel, 2024). Research by Lee, Hwang, and Chen (2022) and Han et al. (2022) found that AI-based chatbots can enhance students' learning attitudes, self-efficacy, and motivation, making students more engaged in the learning process. Peng and Wan (2023) found that students with high social anxiety tend to prefer AI teaching assistants over human ones due to the reduced interpersonal pressure. Klos et al. (2021) observed a significant reduction in anxiety symptoms among students after engaging with AI assistants, suggesting that such tools may be effective in managing anxiety, particularly among university students. Additionally, Mosleh et al. (2024) reported a positive correlation between chatbot use and emotional intelligence (EI) scores, indicating that frequent interaction with AI chatbots could enhance aspects of emotional intelligence. Studente and Ellis (2020) also found that chatbots can help reduce feelings of isolation, especially among freshmen and international students, which can lower dropout rates during the first year of study.

Despite extensive research on the benefits of GenAI in education, there is limited exploration of how students engage in the process of interacting with GenAI tools. To our knowledge, only one published study has investigated the patterns of dialogue between students and generative AI (Han et al., 2023). By analysing the language used in student-ChatGPT conversations, the study concluded that EFL students tend to perceive ChatGPT as a human-like AI, a multilingual entity, and an intelligent peer. Han et al. (2023)'s research offers valuable perspectives on observing and categorising the interaction patterns between students and GPT. However, their study did not specifically analyse the nature of students' questions or explore the connection between the questions and the students' personal backgrounds. Such an analysis is crucial for understanding how GenAI impacts students and the mechanisms through which it operates. Although previous research emphasised GenAI's effectiveness in enhancing academic performance, few studies examined the underlying mechanisms driving these improvements. The "how" and "why" behind GenAI's impact on learning outcomes are not yet fully understood.

    Understanding how interactions occur is a crucial prerequisite for exploring why and how generative AI has an impact. Therefore, this study aims to gain in-depth insights into the nature of the student-AI interactions and students' learning experiences through a multi-faceted approach, including the analysis of student-AI

interaction transcripts, questionnaires, and interviews. Three main questions are investigated in this study:

1. How do students interact with the GenAI teaching assistant?
2. Do different students exhibit distinct interaction patterns?
3. How do students perceive their interactions with the GenAI teaching assistant?

## 3. Methodology

### 3.1 Participants and settings

The participants of this study consisted of 20 students from the "Advanced Quantitative Methods in Educational Research" course at an elite university in China. This course focused on the design, implementation, and interpretation of advanced quantitative methods used to evaluate causal relationships in educational research. Of the 20 participants, 40% (N=8) were female, and 60% (N=12) were male. The mean age of all participants was 23.95 (SD=1.19). The participants primarily consisted of master's students, with the addition of 4 doctoral students. The study received ethical approval from [NAME University] (name hidden for anonymous peer review). Informed consent was obtained from all participants who agreed to participate in the study.

During the 10-week duration of the course, students interacted with a GenAI teaching assistant specifically designed for this course. The assistant was powered by GPT-4.0, and it was built upon the integration of a core knowledge base and an attribute knowledge base. The core knowledge base was developed using previous course materials, slides, textbooks, and historical Q&A records between students and instructors. The attribute knowledge base contained essential course-specific information, instructor details, and role-specific data tailored for the GenAI assistant. This dual knowledge base ensures that the platform performs excellently in both role positioning and providing expert-level responses. The AI assistant interface is shown in Figure 1.

**Figure 1**

*The interface of the GenAI teaching assistant*

To check the accuracy of the GenAI assistant's responses, the course instructor and a human teaching assistant conducted a preliminary trial. They engaged in over 50 rounds of interaction with the AI assistant based on frequently asked questions from previous courses. The accuracy rate of the AI assistant's responses exceeded 95%. To ensure that students receive accurate responses from the AI teaching assistant during the course, the instructor and human teaching assistants reviewed the dialogues between students and the AI assistant every week, verifying the accuracy and reliability of every response provided by the AI. For any problematic answers given by the AI, the instructor and human teaching assistants annotated the response and used the platform's built-in automatic feedback mechanism to remind students. Additionally, the instructor also provided direct feedback in the classroom to alert students to any corrections.

**3.2 Study design**

This study employed a mixed-methods case study approach, chosen for its suitability in providing an in-depth exploration of student learning experience with a GenAI-based teaching assistant within the real-world context of a 10-week course. The case study method allowed for a detailed examination of the processes and outcomes of these interactions within a bounded educational system, while the mixed-methods design was utilized to capture the multifaceted nature of these interactions and can provide a comprehensive understanding of both the nature of student-GenAI interaction and student perceptions. Both quantitative methods and qualitative methods were used, including surveys, interaction records, and semi-structured interviews. This design enabled us to map the distribution of AI usage throughout the course, categorise the types of questions students asked, explore the relationship between students' backgrounds and their usage patterns, and delve into students' perceptions of the benefits and drawbacks of GenAI in supporting their learning.

### 3.3 Data collection

This study employed both quantitative and qualitative data. The quantitative data comprised survey results. The qualitative data included students' dialogue with the AI assistant and interview transcripts. The data collection methods are detailed as follows.

#### 3.3.1 Survey

*Mathematical background*

We surveyed students' mathematical backgrounds, including the number of calculus, linear algebra, probability and statistics, and econometrics courses they had completed. Based on the number of relevant courses taken, we categorised students into high mathematical foundation (N=10) and low mathematical foundation (N=10) groups.

*Attitudes towards the GenAI teaching assistant*

As mentioned earlier, students' use and perceptions of the GenAI teaching assistant were examined via the survey after the course. Students were asked to rate the frequency of usage of the GenAI teaching assistant, their reliance on the GenAI teaching assistant, the information accuracy and reliability of the GenAI assistant, compared with traditional learning approaches such as the human teacher's instruction, human teaching assistant's Q&A, course materials, and Moocs or recorded course lectures.

#### 3.3.2 Records of student dialogue with AI assistant

All students' dialogues with the GenAI assistant were collected, which included the user ID, the time each message was sent, the number of tokens in each message, and the content of the message. These data were then subjected to coding analysis by coders.

### 3.3.3 Interviews

During the later stages of the course, the researcher conducted ten semi-structured interviews, each lasting approximately 30-40 minutes, focusing on students' learning experiences and their attitudes toward the GenAI assistant. These interviews were audio-recorded and then transcribed into text for analysis.

## 3.4 Data analysis

### 3.4.1 Quantitative analysis

Descriptive analysis was firstly employed to examine the students' use of the AI teaching assistant. The frequency of student interactions with the AI assistant was analysed. The different AI assistant usage among students with varying levels of mathematical proficiency was analysed through independent t-tests and Poisson regression.

### 3.4.2 Analysis of student inquiries

In this study, students' interaction with the AI teaching assistant was analysed from two primary dimensions: the cognitive level of students' questions based on Bloom's Taxonomy and the question formulation characteristics based on Lo (2023)'s CLEAR framework. The two frameworks served as tools for the coding scheme, and student inquiries were coded manually. Three researchers attended a pre-coding training session to discuss the codebook and conduct a coding trial. All the disagreements in the trials were discussed and solved. Subsequently, two researchers independently coded all student inquiries. Daily discussions were held to address any disagreements that arose during the coding process. When the two coders could not reach a consensus, the third researcher was invited to join the discussion and made a decision. The inter-rater reliability was 0.89.

*Bloom's Taxonomy*

The cognitive levels of students' questions were primarily characterised based on Bloom's Taxonomy of Cognitive Domains. Bloom et al. (1956) developed a widely accepted taxonomy for categorising educational goals and cognitive processes. It organises learning into six levels, from lower-order thinking skills (remembering, understanding, applying) to higher-order skills (analysing, evaluating, creating) (Anderson & Krathwohl, 2001). The taxonomy is widely used in educational contexts to assess student learning outcomes and design instructional activities (Chandio et al., 2016). Crompton et al. (2018) noted that Bloom's Taxonomy has also been applied to structure learning activities and analyse student inputs in AI-enhanced learning environments. Using Bloom's Taxonomy to code student questions to the GenAI assistant can provide insights into their cognitive engagement and depth of understanding. By examining these questions, we can infer students' cognitive positioning within the taxonomy.

*CLEAR framework*

The question formulation is primarily characterised based on Lo (2023)'s CLEAR framework. Lo (2023) introduces the CLEAR Framework, an essential tool for enhancing students' critical thinking and information literacy skills. Structured around five core principles — Concise, Logical, Explicit, Adaptive, and Reflective — this framework guides educators in creating effective prompts that optimise the quality of AI-generated responses, thereby improving learning and information retrieval processes. This framework was selected for its ability to evaluate the cognitive and communicative aspects of student queries.

### 3.4.3 Analysis of interview data

For the analysis of the interview data, a thematic analysis approach was employed. This method allows the researcher to systematically identify, organise, and interpret recurring themes within the transcribed interviews. The researcher first read through the transcriptions multiple times to gain an overall understanding of the data. After this, initial codes were generated based on significant patterns and key points expressed by the students regarding their experiences and attitudes toward the AI assistant. Substantially, codes were grouped into broader themes, which encapsulated the core aspects of the students' learning experiences. Themes such as engagement with AI, perceived benefits and challenges, and impact on learning outcomes emerged. These themes were then reviewed and refined to ensure they accurately reflected the data.

## 4. Results

### 4.1 Student use of GenAI teaching assistant

Over the ten-week period, nineteen students engaged in 1,418 rounds of dialogue with the AI tutor. Among these, 61 rounds of dialogue were deemed irrelevant to the course, for example, students asked, "Who are you?" and "Can I ask you about qualitative research?". Since this AI agent was trained based on a quantitative research methods course, its responses to non-course-related questions were brief and basic. In this study, we analysed the 1,357 rounds of dialogue that focused on the course content.

The bar chart of Figure 2 presents the weekly token count of student inputs over the ten weeks. It is evident that student enthusiasm for using the AI tutor was relatively high at the beginning of the course. However, there was a noticeable decline in usage during the middle weeks. Particularly, student usage was very sparse in Week 7, probably because of the public holidays.

**Figure 2**

*Distribution of tokens over 10 weeks*

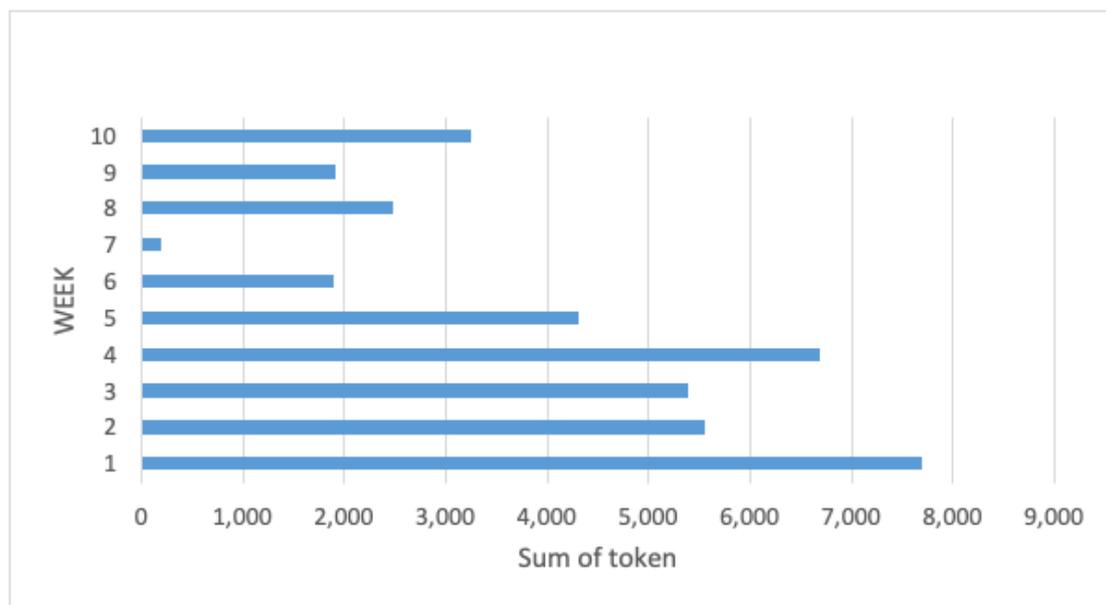

We compared the average number of messages sent by students with different levels of mathematical foundation using an independent t-test and found that students with stronger mathematical foundations significantly used the AI teaching assistant less frequently ($p < .001$). On average, they sent 43.5 messages over the 10-week period, among which only 1.8 messages on average were follow-up questions, while students with weaker mathematical foundations sent significantly more messages, with an average of 89.25 messages and 16.5 of which were follow-up questions.

**Figure 3**

*Average Number of Messages Sent by Students with Different Levels of Mathematical Foundation*

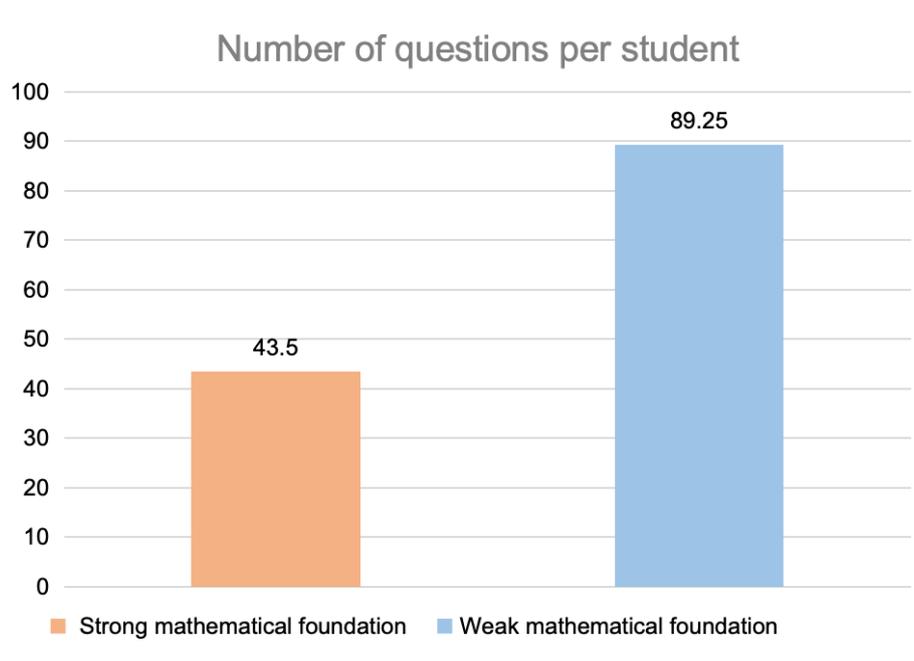

### 4.2 The type of student questions

We classified the 1,357 student messages sent to the AI agent into questions (N=1,245) and other types of expressions (N=112). Among the 1,245 questions, 71 questions asked how to solve homework questions. Although students were advised not to directly ask the AI tutor about specific homework questions, some still did so. For the remaining 1,174 questions, we categorised them based on Bloom's taxonomy. The majority of student questions pertain to knowledge (29%), comprehension (36%), and application (27%), while questions reflecting higher-order thinking, such as analysis (5%), evaluation (2%), and creation (1%), are notably rare. Examples of these questions are listed in Table 1.

**Table 1**

*Categorisation of student questions*

| Category | Examples | N | Percentage |
|---|---|---|---|
| **Knowledge** | *What is asymptotic unbiasedness?* | 341 | 29% |
| **Comprehension** | *Why is the variance of the residuals equal to the expected value of the squared residuals?* | 423 | 36% |
| **Application** | *Analyse mediation and moderation effects using Stata* | 316 | 27% |
| **Analysis** | *Does the above result imply that the lag effect k=1 is not significant, but k=2 is significant?* | 59 | 5% |
| **Synthesis** | *How should interrupted time series analysis be designed to address the three questions raised above?* | 23 | 2% |
| **Evaluation** | *Is it appropriate to use the aforementioned DiD model without a control group, meaning without middle schools that did not participate in the teaching evaluation?* | 12 | 1% |
| | | 1174 | 100% |

In addition, there were 112 rounds of dialogue that were not directly related to the course content and thus were not categorised using Bloom's taxonomy. Among these dialogues, 36 requests were about translation. Since all course materials and assignments were in English, some students asked the AI agent to translate some course materials or literature from English to Chinese. Additionally, 28 dialogues affirmed the AI-generated content, exemplified by expressions like "Thanks.", "You answered very well!", and "Now I understand, it's astonishing!". Furthermore, 21 dialogues conveyed noticeable dissatisfaction or negative sentiment towards the generated responses, often prompting requests for modification or clarification. Examples include statements such as "Isn't this incorrect?" and "I'm a bit confused.". There were also inquiries about how to use the AI tutor or requests for continued content generation, such as "How do I share a file with you?", "Continue answering" or "Please answer again".

Heatmaps below show the distribution of question categories for students with a higher mathematical foundation and those with a lower mathematical foundation over the ten weeks. We used Poisson regression to analyse the frequency and types of questions asked by students with different levels of mathematical foundation. The results indicate that students with a weaker mathematical foundation asked significantly more questions than those with a stronger mathematical foundation. Moreover, the number of knowledge questions was significantly higher than other questions. Additionally, student usage was significantly lower in the latter five weeks compared to the first five weeks. Regarding the interaction effects, overall, more application-type questions and a greater total number of questions reflecting higher-order thinking skills were asked in the latter weeks. Students with a weaker mathematical foundation asked significantly more questions regarding remembering and questions reflecting higher-order thinking than those students with. However, compared with students with a higher mathematical foundation, students with a lower mathematical foundation asked significantly fewer complex questions in the latter five weeks.

**Figure 4**

*Distribution of question types among different groups of students across 10 weeks*

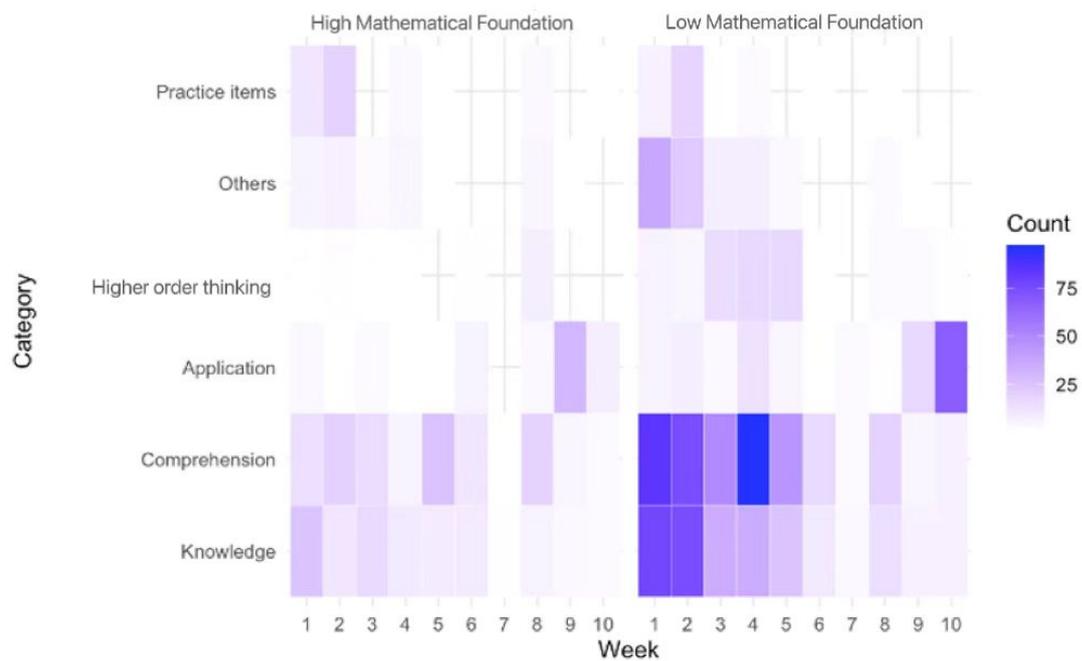

Note: Week 7 coincided with a public holiday in China.

### 4.3 The nature of student expressions

To better understand the characteristics of students' questions, we rated the conciseness, logic, explicitness, adaptability, and reflectiveness of each question. Then we employed ANOVA models to compare the characteristics of different types of questions. The results are illustrated in Table 2. Regarding the conciseness, knowledge-based questions were rated significantly higher than the other four types (*p*s < .001). Comprehension questions were also significantly more concise than application questions and questions reflecting higher-order thinking skills (*p*s< .001). Besides, the students' questions about knowledge comprehension are expressed in the most logical and explicit manner, significantly higher than conceptual questions (*p*< .001) and application questions (*p*= .002, *p*= .001, respectively). Additionally, questions that reflected higher-order thinking were the most adaptive and reflective, scoring significantly higher compared to other types of questions (*p*< .001). Knowledge questions scored the lowest on adaptability and reflectiveness, significantly lower than the other types of questions (*p*< .001).

**Table 2**

*Ratings Based on the CLEAR Framework for Different Question Categories*

|  | Knowledge | | Comprehension | | Application | | Complex | | F | p | η² |
|---|---|---|---|---|---|---|---|---|---|---|---|
|  | M | SD | M | SD | M | SD | M | SD | | | |
| **Concise** | 4.96 | 0.21 | 4.72 | 0.5 | 4.46 | 0.89 | 4.29 | 0.65 | 60.56 | <.001 | 0.135 |
| **Logical** | 3.74 | 1.77 | 4.63 | 0.73 | 4,24 | 1.23 | 4.46 | 0.77 | 38.74 | <.001 | 0.091 |
| **Explicit** | 3.75 | 1.77 | 4.6 | 0.77 | 4.17 | 1.23 | 4.39 | 0.85 | 34.91 | <.001 | 0.083 |
| **Adaptive** | 1.84 | 0.7 | 2.57 | 0.79 | 2.23 | 0.78 | 3.55 | 0.8 | 143.81 | <.001 | 0.271 |
| **Reflective** | 1.41 | 0.74 | 2.74 | 0.88 | 2.09 | 0.99 | 3.85 | 0.7 | 286.23 | <.001 | 0.425 |

We also compared the characteristics of the initial questions and follow-up questions using independent t-tests. As shown in Table 3, follow-up questions scored lower on the "concise" dimension (t=-6.89, *p*< .001) but higher on the "adaptive" and "reflective" dimensions (t=18.99, p < .001 and t = 13.41, p < .001, respectively).

**Table 3**

*Ratings Based on the CLEAR Framework for Initial and Follow-up Questions*

|            | Initial Questions | | Follow-up questions | | t(19) | p |
|------------|------|------|------|------|--------|-------|
|            | **M** | **SD** | **M** | **SD** | | |
| **Concise**   | 4.78 | 0.53 | 4.48 | 0.66 | 6.88   | <.001 |
| **Logical**   | 4.25 | 1.36 | 4.34 | 0.89 | -0.85  | 0.199 |
| **Explicit**  | 4.23 | 1.37 | 4.3  | 0.89 | -0.65  | 0.26  |
| **Adaptive**  | 2.16 | 0.74 | 3.32 | 0.92 | -18.99 | <.001 |
| **Reflective**| 2.1  | 1.05 | 3.2  | 0.96 | -13.41 | <.001 |

  Additionally, we compared the characteristics of questions posed by students with strong and weak mathematical foundations. Questions from students with a strong mathematical foundation scored significantly higher on the "logical" and "explicit" dimensions compared to those from students with a weak mathematical foundation. However, in the "reflective" dimension, students with a weak mathematical foundation scored significantly higher, probably because they asked more follow-up questions.

**Table 4**

*Ratings Based on the CLEAR Framework for Students with Strong and Weak Mathematical Foundations*

|            | Strong Mathematical | | Weak Mathematical | | t(19) | p |
|------------|------|------|------|------|-------|-------|
|            | **M** | **SD** | **M** | **SD** | | |
| **Concise**   | 4.75 | 0.52 | 4.72 | 0.57 | 0.92  | 0.18  |
| **Logical**   | 4.53 | 1.03 | 4.2  | 1.35 | 4.69  | <.001 |
| **Explicit**  | 4.53 | 1    | 4.17 | 1.37 | 5.03  | <.001 |
| **Adaptive**  | 2.34 | 0.78 | 2.36 | 0.92 | -0.55 | 0.292 |
| **Reflective**| 2.13 | 1.05 | 2.27 | 1.13 | -2.09 | 0.019 |

**4.4 Student perceptions of the use of the GenAI learning assistant**

**4.4.1 Comparison between AI learning assistant and traditional learning methods**

At the end of the semester, we invited students to compare the use of the AI teaching assistant with traditional learning approaches regarding usage frequency, reliance, information accuracy, and reliability. We have conducted a series of paired samples t-tests to compare students' perceptions of the use of the GenAI teaching assistant with those of traditional learning approaches. The results revealed that students reported using the AI tutor significantly more frequently than consulting with teachers (t=3.86, $p$<.001) or human teaching assistants (t=3.71, $p$<.001). Students' reliance on the AI tutor was significantly higher than their reliance on human teaching assistants (t=4.53, $p$<.001), with no significant difference in reliance between the AI teaching assistant and teachers (t=0.00, $p$=.500). Students found the AI teaching assistant's support to be significantly more helpful than that provided by human teaching assistants (t=2.38, $p$=.014), with no significant difference in helpfulness between the AI teaching assistant and the human teacher (t=-0.42, $p$=.340).

Results also showed some reservations about students' trust of the information provided by the AI teaching assistant. Students perceived the accuracy of information provided by AI to be significantly lower than that provided by human teachers (t=--4.66, $p$<.001), textbooks (t=-3.32, $p$=.002), and MOOCs or recorded lectures (t=-2.77, $p$=.006). Similarly, the reliability of information from the AI tutor was rated significantly lower than that from teachers (t=-4.77, $p$<.001), textbooks (t=-2.60, $p$=.009), and MOOCs or recorded lectures (t=-2.77, $p$=.006).

**Figure 5**

*Student perceptions on the GenAI learning assistant and traditional learning methods*

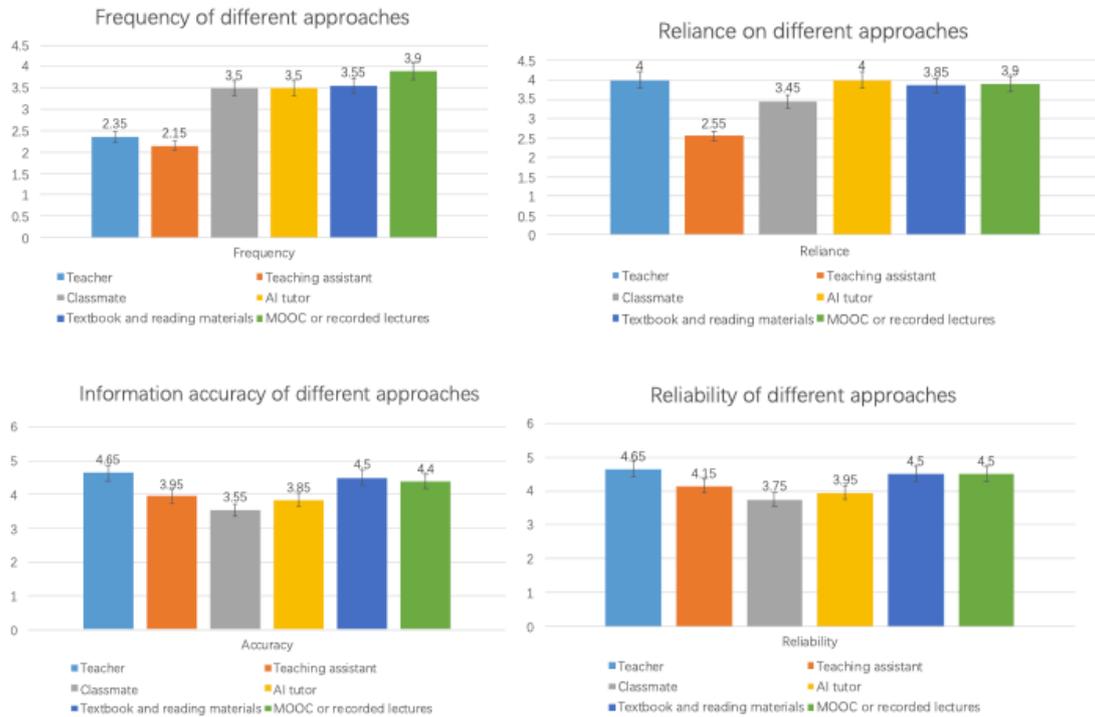

Based on these findings, we conducted further interviews to explore students' learning experiences. During the interviews, we invited students to share in which scenarios they used the AI teaching assistant, how they used it, and their views on its effectiveness.

**4.4.2 Advantages and concerns of the use of the GenAI teaching assistant**

The results revealed three primary scenarios in which students utilised the AI teaching assistant:

*Usage scenarios*

**Course preparation and review:** All students mentioned using the AI teaching assistant during their pre-class preparation and review sessions. They relied on the AI to help them understand challenging reading materials or unfamiliar concepts.

> *"To prepare for the course, I usually read the textbook first, and then I ask the AI tutor how it would explain a specific concept that I am not familiar with. Because the AI might have its ways of describing things, which could help me understand the concept or a theory better."* (S01)

**Assisting with Course Assignments**: Students frequently used the AI teaching assistant for help with coursework, particularly with data analysis tasks using Stata.

They asked the AI about code operations or to explain code errors. Additionally, the AI assisted with literature comprehension and translation tasks.

> *"I use it when I read literature. I asked it to explain things I don't understand in the literature."* (S02)

> *"When writing assignments in English, I write in Chinese first and then ask the AI to translate it. Sometimes I used the AI to translate the English paper to Chinese so I can understand better."* (S07)

**Providing Support During Classes**: Some students used the AI assistant during classes to seek immediate explanations for points they didn't understand in lectures.

> *"I open it during class because sometimes I don't understand what the teacher is saying, so I ask the AI for more explanations right away."* (S03)

*Advantages of the GenAI teaching assistant*

The interviews also highlighted several key advantages of the AI assistant from the students' perspectives. These advantages contributed to a more positive and effective learning experience, addressing some of the challenges they faced with traditional learning methods.

**Timely and Unrestricted Assistance**: One of the most significant benefits reported by students was the AI tutor's availability. Unlike human tutors or professors, the AI tutor was accessible 24/7, allowing students to ask questions and receive answers at any time. This constant availability meant that students could study and complete assignments at their own pace without being constrained by the availability of human assistance. The flexibility to ask unlimited questions without feeling like they were imposing on someone else was particularly valued. A student described it as *"it is tireless and willing to listen to my continuous questions, very professionally."* (S01).

**High Efficiency and Quality of Responses**: Overall, students were satisfied with the AI teaching assistant's efficiency and the quality of its answers. Students mentioned that the AI could quickly generate responses, often within seconds, which was much faster than waiting for a human tutor to reply. The quality of the responses was also noted to be high, with the AI providing structured and comprehensive answers. Students also believed that the AI teaching assistant was adept at delivering the most critical and relevant information directly, without including extraneous details that might distract or overwhelm the students.

> *"Compared to textbooks, the AI assistant can provide a very structured answer directly… The AI assistant integrates and searches out the knowledge you don't know, eliminating unnecessary details found in books and giving you the core and essential point."(S04)*

Some students particularly appreciated the AI tutor's help when they had weak foundational knowledge, as it provided precise and essential assistance. Through multiple interactions with the AI tutor, students could deepen their understanding and gain some new insights.

> *"I ask the AI assistant if my understanding is correct, and it rephrases my thoughts and tells me if I am right or wrong. This interaction is beneficial because it helps me refine my understanding and gives me new ways to analyse or think about the knowledge point."* (S05)

**Reduced Social Pressure**: All students mentioned the reduced interpersonal pressure when using the AI teaching assistant, which was a significant factor in their positive experience. Many students, particularly introverts, mentioned that they could ask questions without feeling self-conscious or worried about how they might be perceived by others. This reduction in social anxiety enabled them to seek help more freely and frequently than they might with a human tutor.

> *"I'm an introvert, so I try to avoid bothering others. I felt more comfortable asking questions to the AI assistant. I can ask silly questions, without experiencing social anxiety or fear of judgement."(S09)*

Even extroverted students acknowledged the convenience of not having to wait for responses or decide who to approach.

> *"As an extrovert, I don't mind asking people questions, but I still have to wait for their response and choose the right person to ask. The AI tutor offers an opportunity to get an immediate, relatively accurate answer without any interpersonal pressure, especially for those objective questions with standard answers."* (S03)

***Limitations of the GenAI teaching assistant***

**Fragmented Learning:** Some students felt that learning with the AI teaching assitant could be fragmented and might have a negative impact on their learning. They shared their concerns that the AI provided piecemeal information rather than a comprehensive, systematic framework of a topic.

> *"Sometimes I ask the AI to explain a concept, but the information is fragmented. When I need a holistic understanding, I still turn to textbooks or other comprehensive sources, like watching recorded lectures online." (S05)*

> *"The efficiency of completing a learning task might improve with the assistance from the AI teaching assistant, but my actual learning might not get improved. What I have learned is how to communicate with the AI rather than a full understanding of the knowledge." (S03)*

**Challenges with Complex Tasks**: students found the AI teaching assistant's responses on complex tasks less satisfactory. One student shared, *"When dealing with complex data models, I had to repeatedly ask the AI assistant and provide detailed descriptions to get useful answers." (S08).* Additionally, complex tasks are often contextual and multidimensional, requiring detailed descriptions. Students feel that describing such tasks in a way that the AI can understand demands high cognitive effort. They find it easier to achieve understanding through real-time conversations with humans.

> "*For complex tasks like completing a statistical table, I preferred asking a classmate who understood the context better than relying on the AI assistant*." (S02)

**Accuracy of information:** Occasional students have encountered inaccurate responses from the AI tutor, making them sceptical about the accuracy of outputs. They felt the need to cross-check the response generated by AI with other sources. One student stated, "*I sometimes doubt the accuracy of the AI teaching assistant's responses and feel the need to verify the information through other reliable and authoritative sources*." (S07) Another student said, *"I often ask the AI to do a self-check or answer the same question multiple times, but I also look up the same information elsewhere to ensure its correctness."* (S09)

One student mentioned the desire to trace the sources of the information provided by the AI, expressing that if the AI could offer its sources, they would trust its output more.

*"Because I don't know where the AI tutor gets its explanations from. I want to trace the source. I often do this: I ask about a piece of knowledge, and it gives me an answer, and then I ask, 'How do you know this? I want to know which websites or places you got this information from'. But it never tells me. I want to know its references, to know whose viewpoint it is citing. I often have this thought, wanting to know the original source of its information. Otherwise, I cannot fully trust the response it provided." (S01)*

## 5. Discussion

### 5.1 Student use of the GenAI teaching assistant

Our study identified several notable patterns in students' usage of the AI teaching assistant, all of which suggest that students may have been engaging with the AI at a relatively superficial level. These patterns include: 1) students interacted with the AI teaching assistant more frequently at the beginning of the course, with a noticeable decline in usage as the course progressed; 2) a significant portion of the inquiries focused on basic knowledge, while more complex or in-depth questions were less common; 3) the frequency of multi-turn dialogues, where students engaged in sustained interactions to explore an issue in depth, was relatively low.

We believe the differences in AI usage between the early and later stages of the course are closely related to the course's progression. In the initial stages, students were likely to ask more questions due to the focus on foundational knowledge during the first half of the course. As students encountered new basic concepts, they may have relied more on the AI teaching assistant to clarify and reinforce their understanding of knowledge. As the course advanced and the tasks became more focused on software operations and comprehensive problem-solving, students' interactions with the AI assistant declined. This phenomenon corresponds to the nature of students' questions, which were predominantly basic and foundational questions aligned with lower-order cognitive skills. Previous research indicated that students capable of formulating high-quality questions demonstrate advanced cognitive abilities, and the complexity of these inquiries reflects the depth of students' thought processes, suggesting that more advanced questions tend to foster deeper cognitive engagement (Bates et al., 2014; Savage, 1998).

These patterns can be explained by two key factors: students' insufficient AI literacy and their understanding of the AI's capabilities. Based on the coding analysis of the nature of student questions, as well as insights from the interviews, we hypothesise that many students may not yet possess the skills to effectively formulate higher-order questions. Their ability to prompt the AI with more complex queries seems to be underdeveloped, as evidenced by the relatively low levels of adaptability and reflectiveness in their inquiries. These low levels of adaptive and reflective questioning may suggest a lack of competence or familiarity with how to fully leverage the AI's potential to support deeper learning. Additionally, the proportion of multi-turn dialogues—where students engaged deeply with a single issue—was minimal. Follow-up questions tend to be more reflective and adaptive, but this requires students to effectively process and engage with the AI's responses, then further craft effective prompts to foster more meaningful interactions with AI. Apparently, this might be a challenge for many students.

Furthermore, previous research has confirmed that perceived usefulness and perceived ease of use likely influenced students' behaviour (Algerafi et al., 2023; Romero-Rodríguez et al., 2023). Our interview data revealed that students generally perceived the AI as more effective in providing answers to basic concept definitions and explanations. Conversely, interviewees were sceptical about the AI's ability to effectively handle more complex, nuanced questions. This perception may lead to their pragmatic approach to maximise the perceived utility of the AI tool by asking basic, well-defined problems, limiting their engagement with the AI on a deeper level. It is a reflection of students' strategic decisions influenced by their perceptions of AI's strengths and limitations. Future research could explore strategies to enhance AI's capability in supporting higher-order cognitive skills, potentially fostering more complex and in-depth student inquiries.

**5.2 Relationship between student-AI interaction and students' prior knowledge**

Our analysis suggests a relation between students' usage of the AI assistant and their prior knowledge levels. Students with weaker mathematical foundations tended to use the AI assistant more frequently than those with stronger foundations. Interestingly, despite their lower prior knowledge, these students not only asked more questions but also posed more complex and reflective inquiries compared to their peers with stronger mathematical backgrounds. This finding highlights the significant potential of AI assistants to support students with weaker foundations. Previous research has also demonstrated the effectiveness of AI in providing tailored support to lower-performing individuals. This has been validated in other fields such as Law and Consult. For instance, Choi et al. (2024) studied the effect of AI assistance on human legal analysis and found that the lowest-skilled law students achieved the largest improvements. Similarly, a study on consulting tasks has shown that the use of GPT-4 can improve the performance of below-average consultants by 43% and that of above-average consultants by 17%, relative to their own baseline scores (Dell'Acqua et al., 2024). These findings suggest that GenAI has significant potential to act as a lever for narrowing performance gaps.

It is also important to note that, while students with higher mathematical foundations asked fewer questions, their inquiries were more logically structured and explicit. This suggests that prior knowledge plays a key role in shaping the quality of prompts. The stronger the foundation in a given field, the more precise and targeted the questions, which are more likely to stimulate deeper learning. This finding underscores the importance of foundational knowledge in influencing the quality of human-AI interaction and collaboration.

**5.3 Student attitude towards the AI teaching assistant**

Overall, students in this study were actively engaged with the AI teaching assistant and recognised its effectiveness. Many students found the AI assistant to be comparable to, or even significantly better than, human teaching assistants in terms of helpfulness, accuracy, and reliability of the information provided. It is worth noting that in this course, the human teaching assistants were senior PhD students. However, students perceived the materials, information, and feedback provided by the instructor and MOOC platforms as more reliable than those from the AI assistant. This suggests that while the AI assistant was highly valued for its immediate support, students still placed greater trust in traditional resources, likely due to the perceived authority and expertise of the instructor and structured educational content. This preference aligns with previous findings that service attitudes—encompassing the friendliness, enthusiasm, patience, and respect demonstrated by AI agents—can foster users' trust in the service provider (Kuo et al., 2012; Peng & Wan, 2023).

Another notable finding from the interviews was that students frequently mentioned that communicating with the AI agent was private and judgement-free, enabling them to ask "silly" questions without fear of embarrassment. This finding has been widely confirmed by previous research that students felt psychologically safe and more unrestrained when interacting with AI (Peng & Wan, 2024). In contrast, interactions with human instructors and teaching assistants often involved an unspoken pressure to maintain a certain level of competence or image. This finding echoes previous research indicating that students prefer to seek help from computer systems rather than teachers (Lee et al., 2022), probably due to students' perception of teachers as authority figures in traditional East Asian culture (Littlewood, 2000), where the concept of "尊师重道" (respect for teachers and their teaching) is deeply ingrained (Stowell et al., 2010). Students may feel more concerned about their self-presentation in front of teachers. In contrast, when interacting with the AI assistant, the perceived social pressure present in traditional educational settings seemed to be reduced. This lack of pressure enabled students to engage more and seek help without fear of judgment. By eliminating this barrier, the AI assistant appeared to create a more inclusive and supportive learning environment, encouraging students to ask questions they might otherwise withhold.

**Limitations**

Despite the valuable insights gained from this study, several limitations should be acknowledged. This study was conducted within the context of a single course, which focused on advanced quantitative research methods, which may limit the generalisability of the findings to other courses or disciplines, especially those with varying cognitive demands or instructional approaches. Moreover, the relatively small number of students in this course calls for caution when considering the practical applicability of the results. Future research should expand to multiple courses and contexts involving diverse student populations to identify different interaction patterns, thereby gaining a deeper understanding of student-AI interactions and providing more targeted guidance to students.

**Conclusion**

This study explored the nature of student interactions with an AI teaching assistant and examined how these interactions varied based on students' prior knowledge. The findings revealed that students engaged more frequently with the AI assistant in the earlier stages of the course, with most inquiries focused on basic knowledge, while deeper, multi-turn dialogues were less common. Despite this, students with weaker mathematical foundations used the AI more frequently and asked more complex, reflective questions, suggesting that AI has the potential to provide meaningful support for students with lower prior knowledge. Students generally viewed the AI assistant positively, appreciating its accuracy and reliability, often rating it as comparable to, or better than, human teaching assistants. Notably, the AI assistant provided a judgement-free environment that reduced stress and allowed students to ask questions without fear of embarrassment, creating a more inclusive learning atmosphere. However, students still relied more on traditional resources, such as instructors and course materials, for more comprehensive guidance.

In conclusion, while the AI assistant proved to be a valuable educational tool, enhancing students' engagement and supporting their learning, its full potential has yet to be realised. Future research should focus on improving AI literacy among students and designing more rigorous experiments to further investigate the role of AI in addressing the needs of students with diverse learning backgrounds.